# AC calorimetric study of magneto-quantum oscillations in anisotropic multiband $V_2Ga_5$ superconductor

Ye. V. Petrenko[1, 2, 3], J. Kačmarčík[1], T. Samuely[2], J. Haniš[1], S. Gabáni[1],
S. Królak[4], M. J. Winiarski[4], T. Klimczuk[4], M. Gmitra[1] and P. Samuely[1]
[1]*Centre of Low Temperature Physics, Institute of Experimental Physics, Slovak Academy of Sciences, 04001 Košice, Slovak Republic*
[2] *Centre of Low Temperature Physics, Faculty of Science, P. J. Šafárik University, 04001 Košice, Slovak Republic*
[3]*B. Verkin Institute for Low Temperature Physics and Engineering, National Academy of Sciences of Ukraine, Kharkiv 61103, Ukraine*
[4]*Faculty of Applied Physics and Mathematics and Advanced Material Center, Gdansk University of Technology, ul. Narutowicza 11/12, Gdańsk 80–233, Poland*
petrenko@saske.sk

Unlike de Haas–van Alphen measurements, heat-capacity magneto-quantum oscillations directly probe the oscillatory bulk quasiparticle density of states. Here, we report the observation of MQOs in $V_2Ga_5$ single crystals studied via highly sensitive ac calorimetry. The strongest MQO signal is observed for a magnetic field applied along the vanadium chains, in excellent agreement with de Haas–van Alphen magnetization data. A single dominant frequency of 126.6 T resolved by fast Fourier transform confirms the true bulk origin of the elliptical γ Fermi surface pocket located near the *Z* point of the Brillouin zone. The angular dependence of the FFT frequency closely tracks the anisotropy of the γ pocket as supported by first-principles calculations. Analysis of the temperature- and field-dependent MQO amplitudes allows for the precise determination of the effective cyclotron mass, Dingle temperature, quantum relaxation time, carrier mobility, and electron mean free path. Furthermore, we demonstrate that the net Berry flux is invariant with respect to the magnetic field orientation, a consequence of a conserved hybridization phase twist within the γ pocket. These findings establish ac calorimetry as a powerful macroscopic probe of topological orbital hybridization in complex intermetallics.

## Introduction

The V-Ga binary system is known to have superconducting modifications, including $V_3Ga$ with a high critical temperature of 15 K and $V_2Ga_5$ with a lower $T_c$ of ~3.5 K, having strong anisotropy referred to as quasi one-dimensional electronic properties [1–5]. $V_2Ga_5$ has also been shown to feature a multigap superconductivity [4, 6, 7]. Although previous studies have investigated magneto-quantum oscillations (MQOs) in $V_2Ga_5$ using magnetization measurements, thermodynamic evidence from heat capacity remains limited. MQOs through ac calorimetry, efficiently used for studying underdoped $YBa_2Cu_3O$ high-temperature superconductors [8, 9], can provide independent information about the electronic structure and scattering properties of $V_2Ga_5$.

In this work, we investigate the heat-capacity MQOs of $V_2Ga_5$ single crystals using the highly sensitive ac calorimetry technique. We clearly demonstrate the evolution of quantum oscillations as a function of temperature and the relative orientation of the single crystals with respect to the applied external magnetic field. From the experimental data, we extract the effective cyclotron mass, estimate the Dingle temperature, and evaluate the geometric Berry phase. Our measured cyclotron frequencies closely match first-principles calculations of the extremal cross-sectional areas of the γ Fermi pocket centered around the Z point in the Brillouin zone, firmly establishing its true bulk origin. Remarkably, we find a field-orientation-invariant Berry phase close to 0.8π. Given that the γ pocket possesses a trivial band topology, this invariant net Berry flux is shown to arise from the intrinsic V-Ga hybridized states, driven by a hybridization phase twist with a topological winding number of 2π.

## Crystals

Single crystals of $V_2Ga_5$ were grown in $Al_2O_3$ crucibles using V powder (Alfa Aesar, 99.5%) and Ga pieces (Onyxmet, 99.99%), with a 6:94 V:Ga molar ratio. The evacuated quartz tube was heated to 1000 °C at a rate of 100 °C $h^{-1}$, held at this temperature for 48 h, and subsequently cooled to 550 °C at a rate of 2.5 °C $h^{-1}$. At this temperature, the tubes were centrifuged to remove excess Ga flux. The resulting needle-like crystals, with a typical length of 3 mm, were etched in lightly diluted hydrochloric acid (HCl) to eliminate residual Ga from the surface, following the procedure described in [7]. $V_2Ga_5$ crystals were found to be stable in air.

$V_2Ga_5$ crystallizes in the tetragonal space group P4/mbm (No. 127) with a small *c*-value from 2.59 to 2.72 Å and a large a-value from 8.93 to 8.98 Å [1–3]. The crystals naturally grow as needles along the tetragonal [001] direction, coinciding with the V-centered pentagonal channels propagating along the *c*-axis [1–3, 5]. In Fig. 1, we show the crystal structure with a unit cell of the $V_2Ga_5$ [10, 11]. The needle-like morphology along the *c*-axis is important for the present study because the strongest quantum oscillation signal is observed when the magnetic field is applied parallel to this direction.

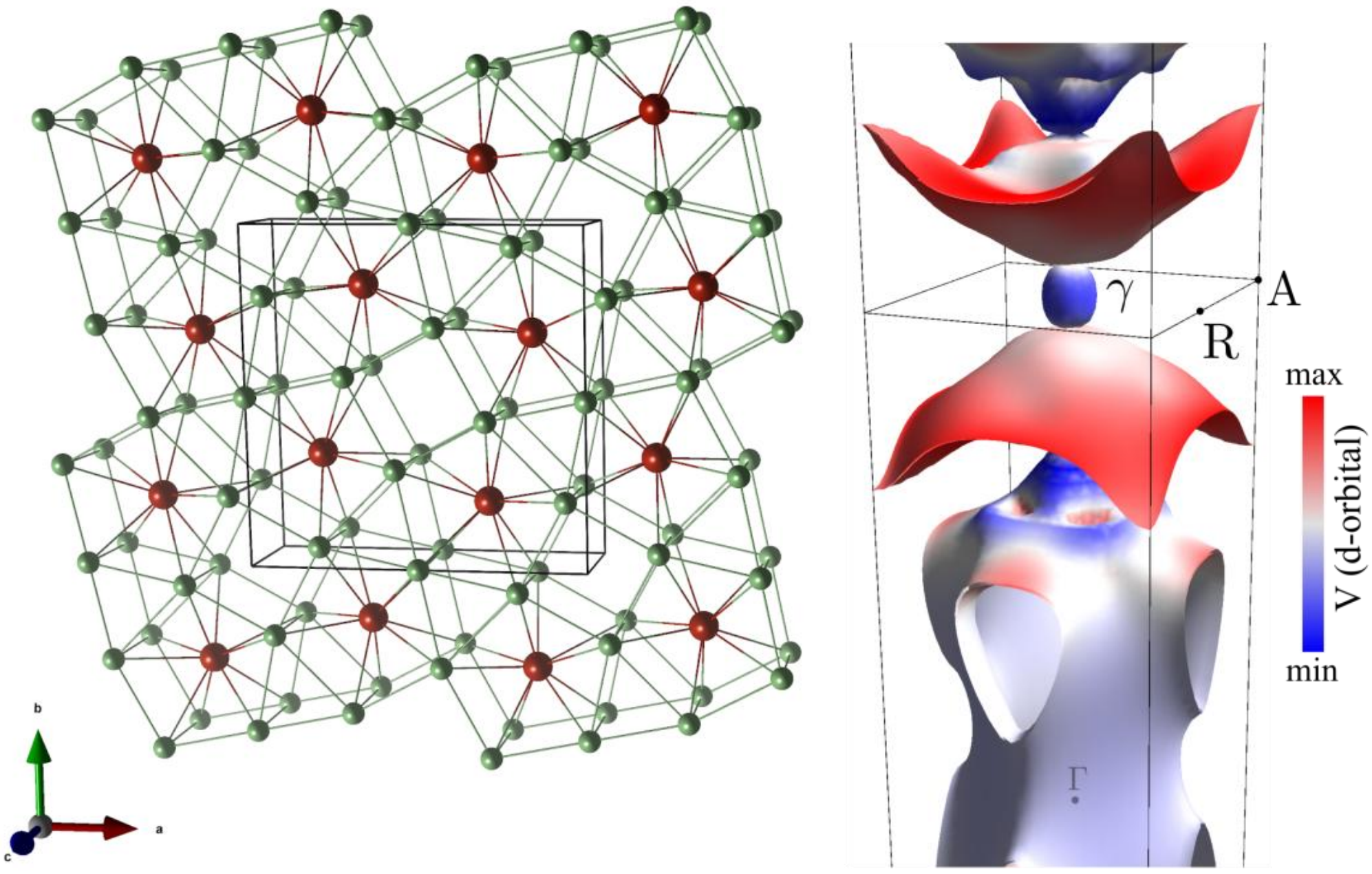

Fig. 1. a) Schematic view of the crystal structure of $V_2Ga_5$. Vanadium atoms are shown with red balls and Gallium atoms with green. The solid box shows one unit cell. b) Calculated Fermi surface with rendered V *d*-orbitals illustrating the γ pocket centered around the Z point.

### Experimental technique

An ac calorimeter installed in the He-3 refrigerator was used to measure the temperature and field sweeps of the heat capacity of the $V_2Ga_5$ single crystalline sample. The technique of ac calorimetry [8, 12–14] employs periodically modulated power on the sample and measures the resulting sinusoidal temperature response. The calorimeter cell consists of a small chip with a resistive thermometer and a heater. The studied sample is attached with low-temperature grease. In our case, heat is supplied to the sample at a frequency of several Hz. In addition to providing the absolute value of heat capacity, the great advantage of ac resistive calorimetry lies in a high sensitivity to relative changes, enabling continuous measurements in minute samples. We performed measurements at temperatures down to 0.3 K and in magnetic fields up to 8 T. In order to obtain the information about MQOs, we fix the selected low temperature and perform a field sweep showing high reproducibility. For a comparison, the magnetization has also been measured at 2 K. These measurements have been performed in the MPMS-3 Quantum Design set up in fields up to 7 T in the vibrating sample magnetometer (VSM) mode.

### Results

#### 1. Observation of the quantum oscillations

For a normal metal, the Lifshitz–Kosevich (LK) theory predicts the magnitude of the quantum oscillations in the heat capacity to be on the order of 0.1% of the electronic specific heat [15–17].

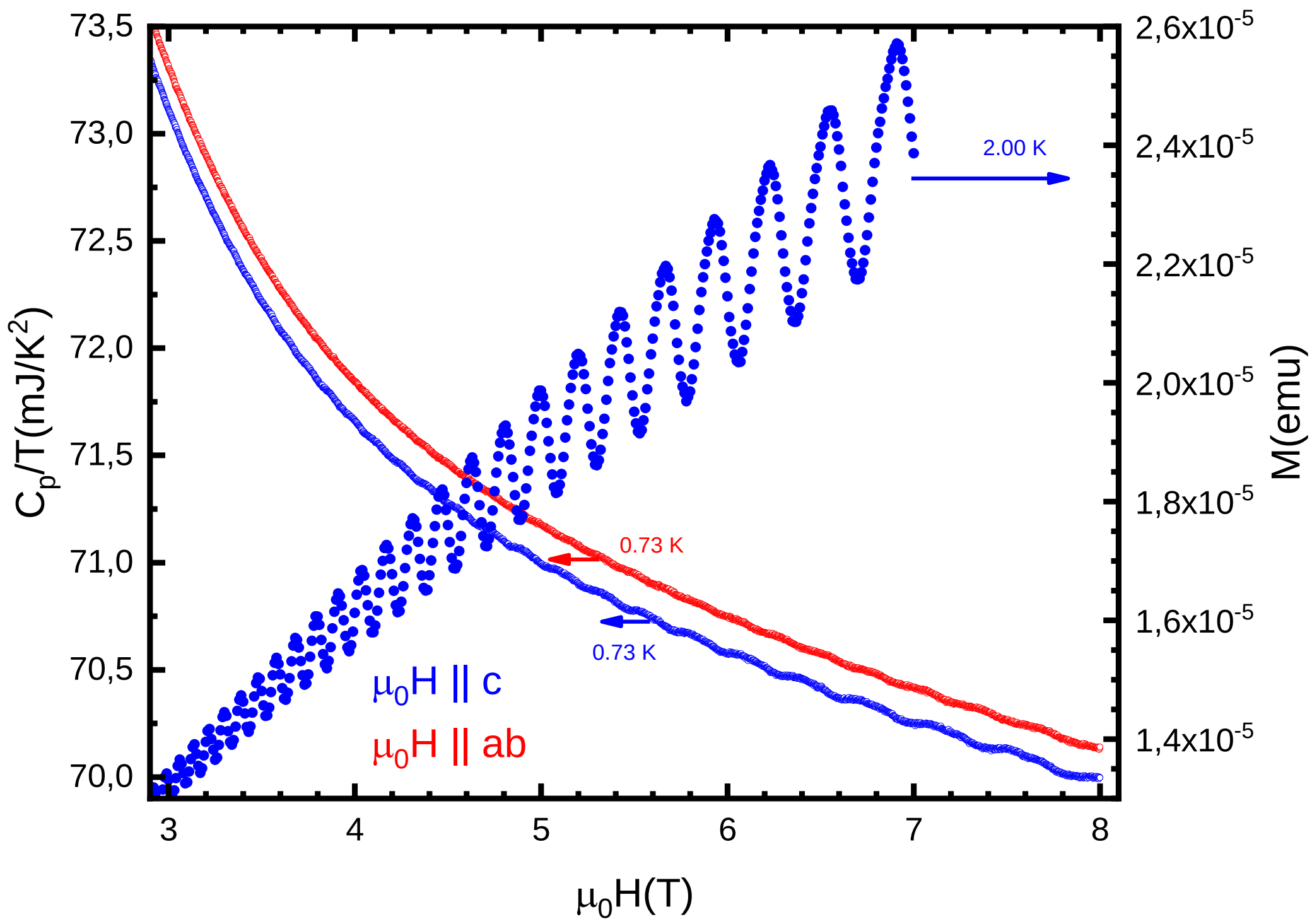

Fig. 2. Quantum oscillations in $V_2Ga_5$ single crystals derived from heat capacity (left axis) and magnetization (right axis). The corresponding color represents the orientation of the sample towards the applied field, blue – $\mu_0 H \parallel c$, red – $\mu_0 H \parallel ab$. The heat capacity MQO is measured at 0.73 K and magnetization MQO at 2 K.

We present the low-temperature heat capacity as well as the magnetization of $V_2Ga_5$ single crystals as a function of applied magnetic field in Fig. 2. Oscillations in the heat capacity are much smaller than those in magnetization but become clearly observable above 3 T. The MQOs of the heat capacity are the most prominent when the field is oriented along the vanadium chain (*c*-axis). The magnetization at 2 K presented in Fig. 2 is measured in a field parallel to the *c*-axis.

Magneto-quantum oscillations arise from oscillations in the thermodynamic potential [15, 16]

$$\tilde{\Omega} = \tilde{\Omega}_0 f_T(z)\,, \tag{1}$$

where $\tilde{\Omega}_0$ is the zero-temperature potential,

$$f_T(z) = \frac{z}{\sinh(z)} \tag{2}$$

is the thermal smearing factor, and

$$z = 14.69\,\frac{pm^*T}{B} \tag{3}$$

is the dimensionless quantity proportional to the electron effective mass $m^*$ (here, it is represented as a normalized value over the bare electron mass $m_e$), temperature, and inverse magnetic field. Here, $p$ is an integer representing the harmonic. The coefficient

$$\frac{2\pi^2 k_B m_e}{e\hbar} \approx 14.69 \tag{4}$$

is the Lifshitz–Kosevich constant with dimensions of Tesla per Kelvin (T/K), where $k_B$ is the Boltzmann constant, $e$ is the electron charge, and $\hbar$ is the reduced Planck's constant.

In the standard Lifshitz–Kosevich model of MQOs for a 3D Fermi surface [16] with extremal area $A_F = \pi k_F^2$ ($k_F$ is the Fermi wavevector), the potential $\tilde{\Omega}_0$, and specific heat (similarly magnetization) are periodic in inverse field $1/B$ with a frequency of oscillation $F$ in Tesla given by

$$F = \frac{\hbar}{2\pi e} A_F \tag{5}$$

The field and temperature dependent amplitude of the oscillatory specific heat corresponding to the $p$-th harmonic of oscillation frequency $F$ can be described according to the LK formula:

$$\frac{C_{osc}(T,B)}{T} = \sum_{p=1}^{N} A_p B^{\frac{1}{2}} p^{-\frac{1}{2}} f_T''(z)\, f_D \cos\left[2\pi p\left(\frac{F}{B} - \frac{1}{2} + \frac{\Phi_B}{2\pi}\right) - \delta\right] \tag{6}$$

where $A_p$ is an amplitude of the $p$-th harmonic, $f_D$ is the Dingle damping term

$$f_D = \exp\left(\frac{-14.69\,pm^*T_D}{B}\right) \tag{7}$$

with the Dingle temperature $T_D$.

The oscillation of $C_{osc}(T,B)$ is described by the cosine term with a phase factor of $-0.5 + \Phi_B/2\pi$, in which $\Phi_B$ is the Berry phase. The phase shift $\delta$ is determined by the dimensionality of the Fermi surface, $\delta = 0$ and $\pm\pi/4$, respectively, for the 2D and 3D cases ($-\pi/4$ for electron-like, and $+\pi/4$ for hole-like cases) [5, 18].

## 2. FFT analysis and oscillation frequency

Prior to frequency analysis, we subtracted a smoothly varying uniform *C(B)*/*T* background obtained using a 9$^{th}$-order polynomial fit at fixed temperatures. Further, by simple linear interpolation, we obtained data sets uniform in *1/B*, which is necessary for the fast Fourier transform (FFT).

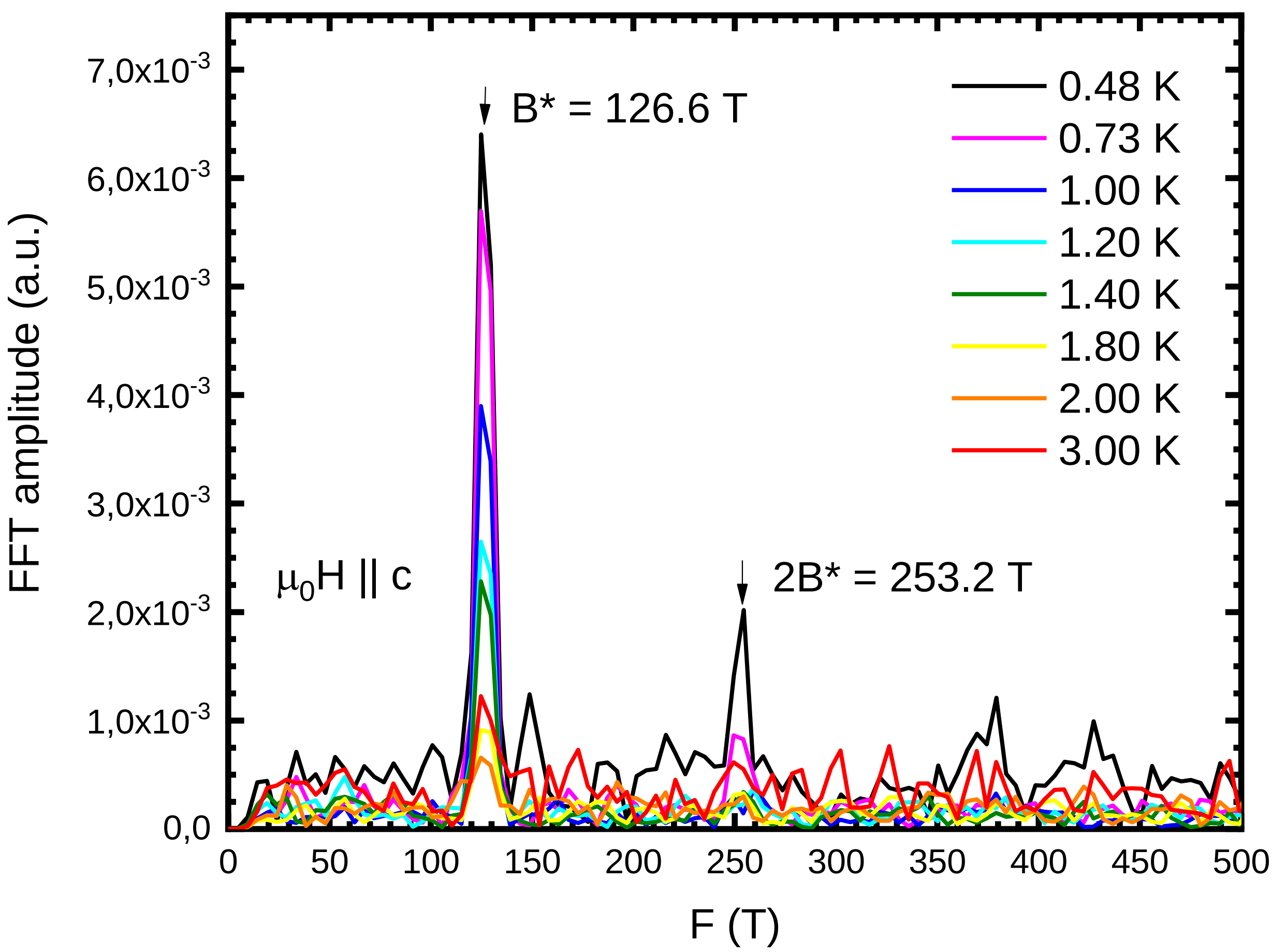

Fig. 3. The FFT spectra of the $C_{osc}/T$ data at different temperatures and fields $\mu_0 H \parallel c$). The peak at 253.2 T represents the second harmonic of oscillations, negligible in further analysis.

By applying the FFT frequency analysis to a field sweep between $B_{min}$ = 3 T and $B_{max}$ = 8 T at selected temperatures we resolve a sharp frequency peak $F$ = 126.6 T with its second harmonic of lower intensity at 253.2 T, shown in Fig. 3. Applying the same FFT analysis to our magnetization measurements yielded the same results. We have not detected any MQO corresponding to the larger magnetic breakdown orbit. We note that in case of $C_{osc}(T,B)/T$, the amplitude of the oscillations allows us to obtain a reasonable signal to noise ratio as a function of inverse magnetic field, though the signal gets smeared much faster at higher temperatures than in the case of magnetization [5].

## 3. Effective mass

There are two ways how to determine the effective mass from the MQOs. The first common approach involves fitting the thermal smearing factor $f_T(z)$ (Eq. 2) to the FFT amplitudes obtained from the $M_{osc}(T, B)$ measurements at different temperatures [5, 18, 19]. However, in the case of heat capacity, we need to apply the $f_T''(z)$ term. And since the FFT peak amplitudes were obtained from the modulus of the Fourier coefficients, the absolute value $|f_T''(z)|$ needs to be considered. Then, using the effective field $B_{eff} = 2 \cdot B_{min} \cdot B_{max}/(B_{min} + B_{max}) = 4.364$ T, we obtain a reasonable fit with $m^* = 0.246m_e$ (Fig. 4).

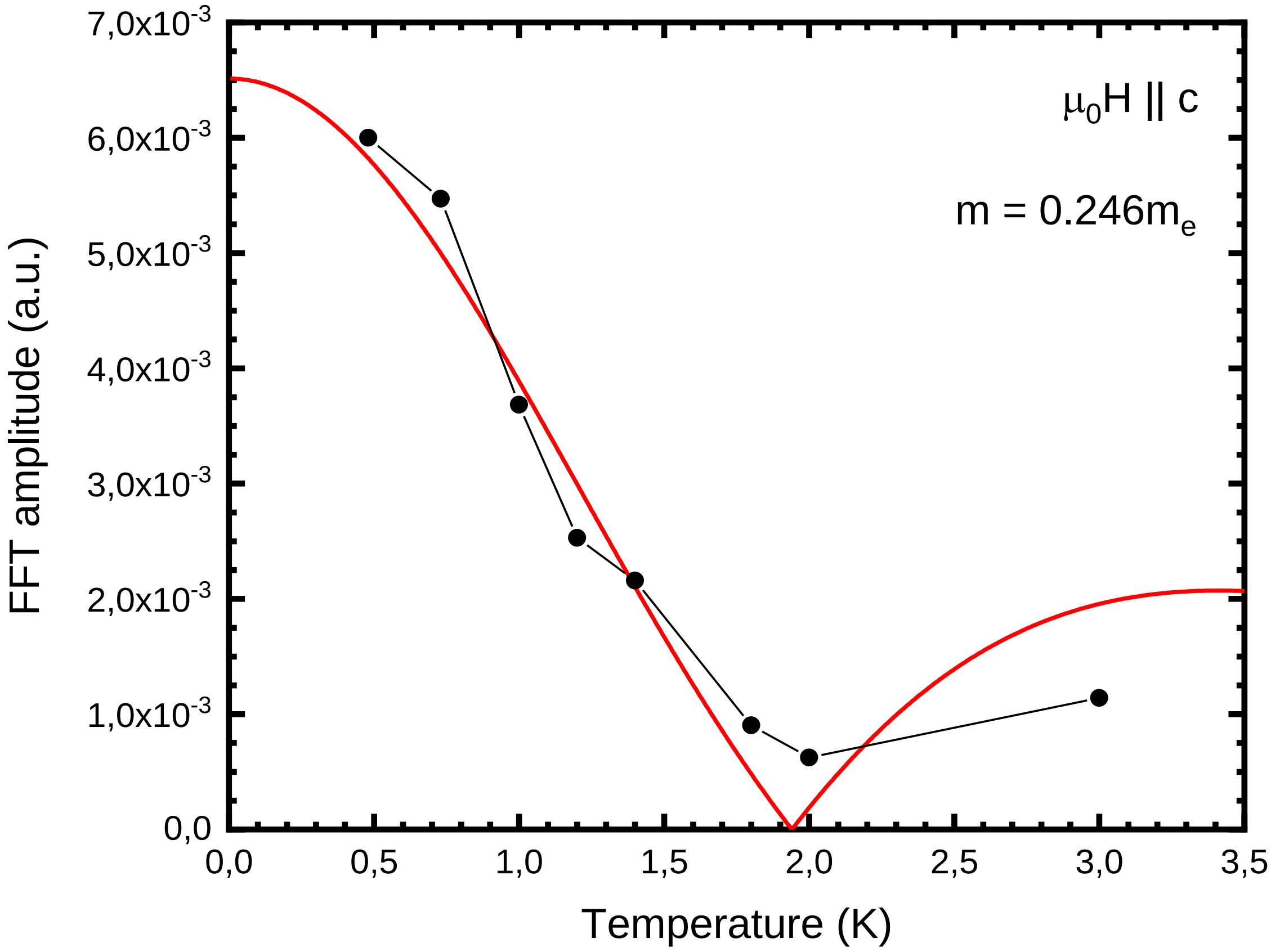

Fig. 4. $|f_T''(z)|$ – fit (red curve) to the FFT amplitude (black dots) for the effective mass determination. A thin black line is a guide for the eyes.

Another approach is to exploit the fact that $f_T``(z_C) = 0$ at $z_C \approx 1.606$. It introduces a node in the amplitude of the oscillatory specific heat as a function of magnetic field, accompanied by a π-phase shift. Since the value of z depends only on the temperature, magnetic field (independent of oscillation frequency), and effective mass [15, 17, 20], it enables us to determine the value of the effective mass from the node's location in field.

In our experiment, we were able to track its position at 4 different temperatures (3.0, 2.0, 1.8, and 1.4 K) within the field window from 3 to 8 T, while observing no features of it below $T_C = 1.2$ K. The latter fact leads to an important preliminary conclusion, that $m^* < 0.273m_e$, since at $z_C$, $m^* = 0.1093B_C/T_C$ as follows from the Eq. 3 for $B_C = 3$ T. The data taken at 2 K and the fit are presented in Fig. 5. Averaging over the values obtained at different temperatures gives $m^* = (0.245 \pm 0.01)m_e$, which means that both methods provide consistent result.

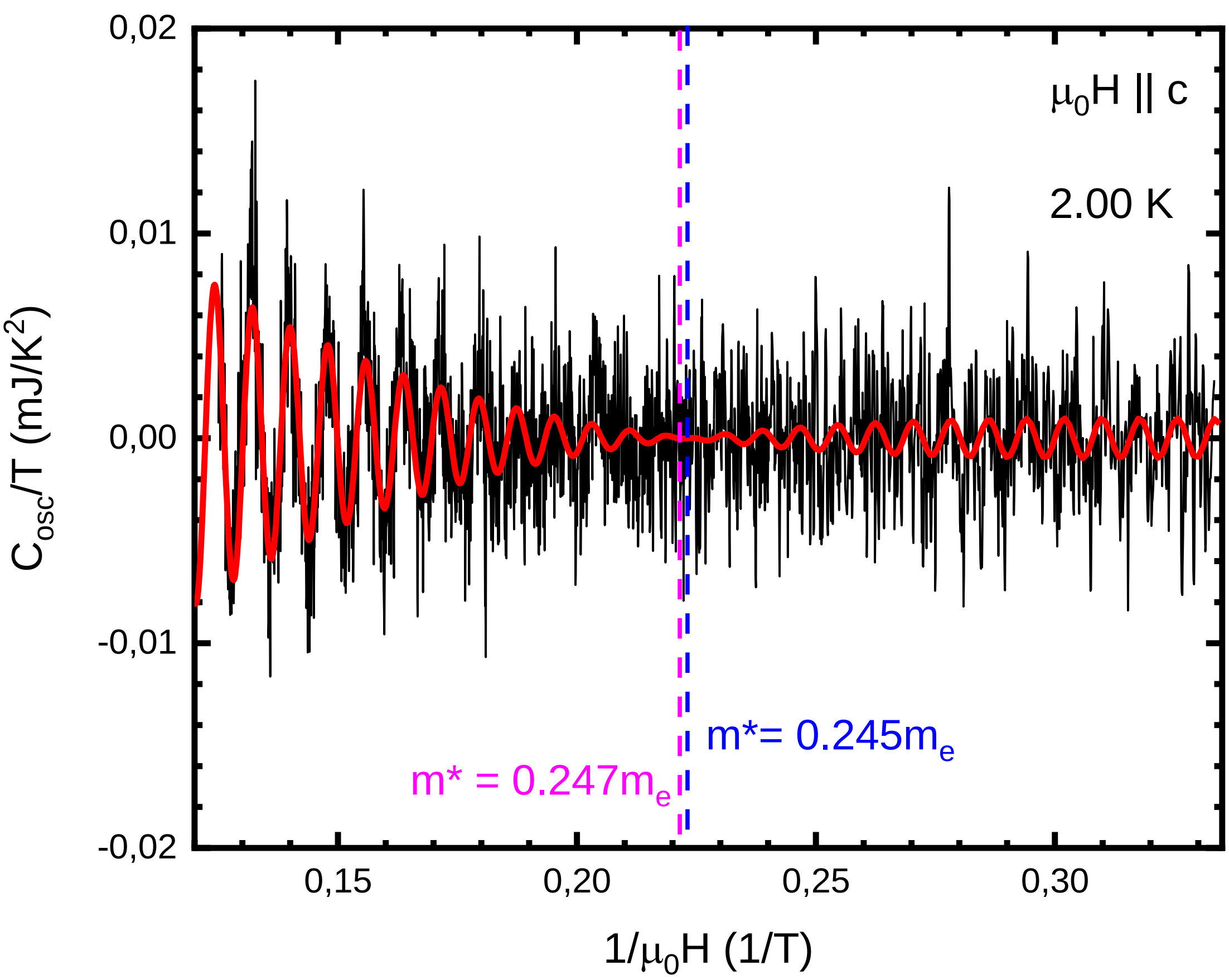

Fig. 5. Experimental $C_{osc}/T$ (black curve, $\mu_0 H \parallel c$) directly fitted to the LK model (red curve) with the only main harmonic of 126.6 T at 2.00 K. The magenta vertical line denotes the node point location, while the blue one shows a position that corresponds to the average $m^*$ value, taking into account measurements at different temperatures.

**4. Dingle temperature**

Another important parameter in the LK model is the Dingle temperature $T_D$. Electron scattering broadens the Landau levels, resulting in an extra amplitude reduction factor, which would have an effect approximately equivalent to that of a rise in temperature. A lower Dingle temperature indicates weaker scattering and sharper Landau levels, which is consistent with the observation of clear quantum oscillations. We fit $C_{osc}(T, B)$ directly at 0.73 K with the highest signal-to-noise ratio (Fig. 6), using $T_D$ and the amplitude of the signal as the only free parameters. The first harmonic was completely satisfactory for the description of all our data available at any given temperature. The calculations give $T_D$ = 1.25 K. Such a low Dingle temperature is reflected in a weak damping of oscillations and suggests low disorder of the investigated single crystals.

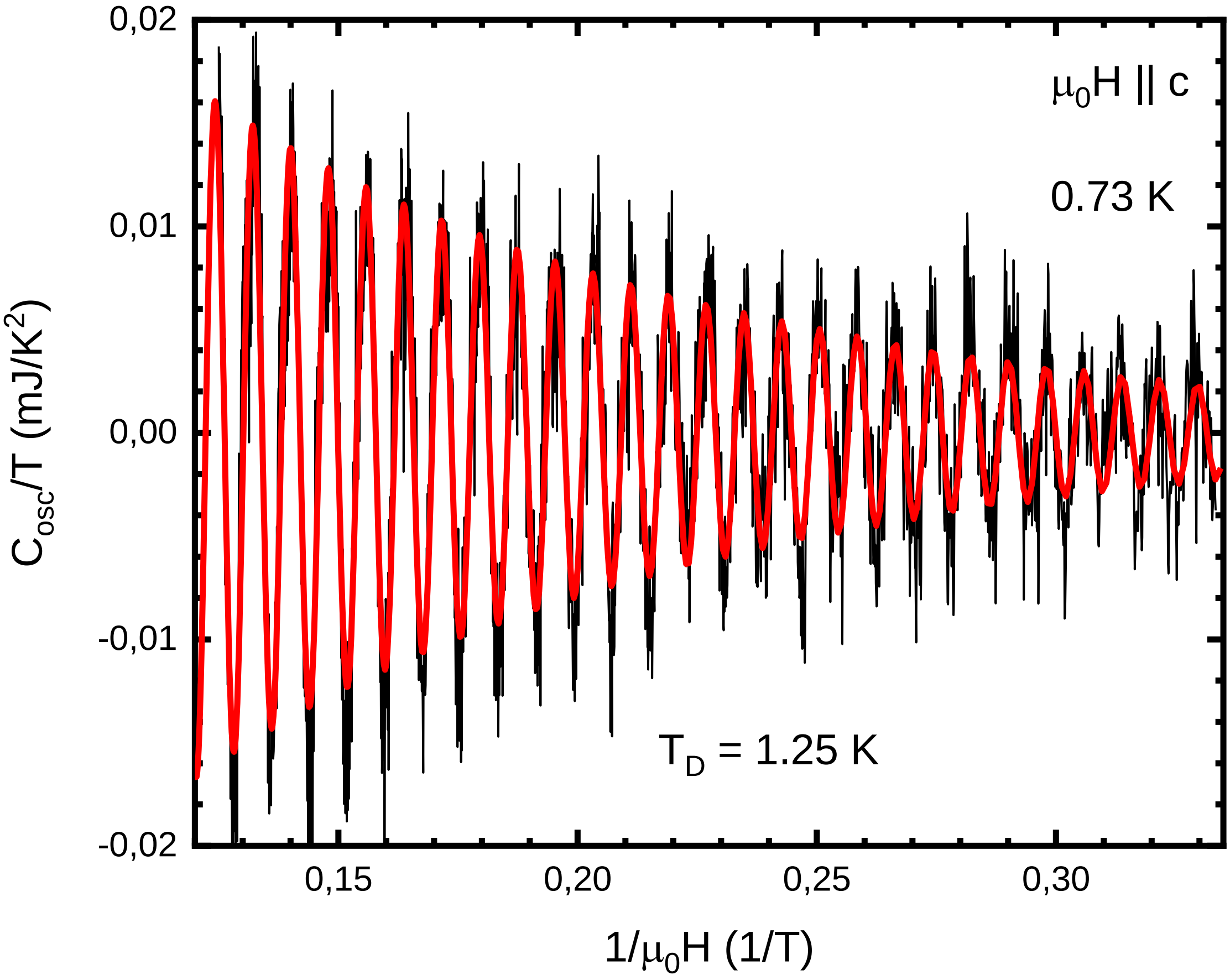

Fig. 6. Experimental $C_{osc}/T$ (black curve, $\mu_0 H \parallel c$) directly fitted (red curve) to the LK model with the only main harmonic of 126.6 T at 0.73 K.

From $T_D$ the quantum relaxation time $\tau = \dfrac{\hbar}{2\pi k_B T_D} = 0.98\cdot10^{-12}$ s and quantum mobility $\mu = \dfrac{e\tau}{m^*}$ = 7010 cm$^2$/(V·s) are calculated. From the free electron model, knowing the relation between $F$ and the extremal area $A_F = \pi k_F^2$ (Eq. 5), we estimate the Fermi velocity $v_F = \dfrac{\hbar k_F}{m^*} = 2.9\cdot10^5$ m/s and the mean free path $l = v_F \tau$ = 287 nm.

Table 1. Parameters derived from quantum oscillations for $V_2Ga_5$. $F$, oscillation frequency; $T_D$, Dingle temperature; $m^*$, effective mass; $\tau$, quantum relaxation time; $\mu$, mobility; $l$, mean free path.

| | $F$ (T) | $m^*/m_e$ | $T_D$ (K) | $\tau$ (ps) | $\mu$ (cm$^2$/(V·s)) | $l$ (nm) |
|---|---|---|---|---|---|---|
| $\mu_0$H ‖ c | 126.6 | 0.245 | 1.25 | 0.98 | 7010 | 287 |
| $\mu_0$H ‖ ab | 174.2 | | | | | |

## 5. Berry phase

The Berry phase $\Phi_B$ was determined independently by directly fitting $C_{osc}(T, B)$ and by constructing the Landau fan diagram by the positions of the chosen kind of extrema [18].

The LK fit of $C_{osc}(T, B)$ at 2 K shown in Fig. 5 yields the $\Phi_B = 0.80\pi$. This value was obtained by taking $\delta = +\pi/4$, which is based on the fact that the prevailing charge carriers in the γ Fermi pocket are holes occupying the ellipsoid pocket with positive curvature [6].

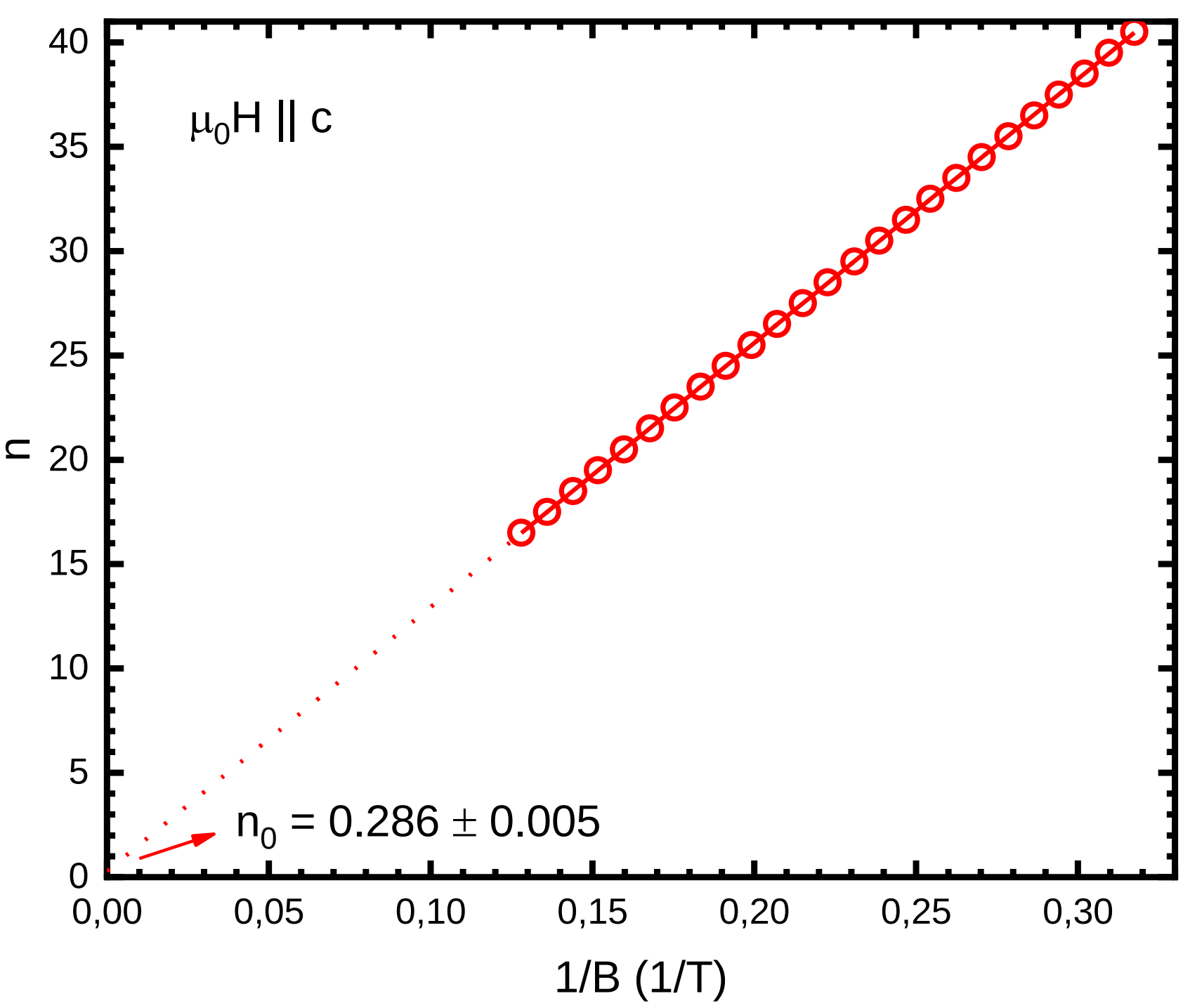

Fig. 7. The Landau fan diagram constructed by minima of $C_{osc}/T$ resulted in the Berry phase of 0.82π (δ = +π/4).

In the Landau fan diagram, the minima of the aforementioned dependence were assigned to the Landau level (LL) indices *n* [5]. The dotted line in Fig. 7 is a linear fit of the LL indices that gives an intercept of $n_0 = 0.286 \pm 0.005$ corresponding to the Berry phase of $\Phi_B = 2\pi|0.286 \pm 0.005 + \delta| = 0.82\pi$ ($\delta = +\pi/4$) while $F = 126.5$ T, which is in perfect agreement with our FFT analysis.

## 6. Angular dependence of quantum oscillations

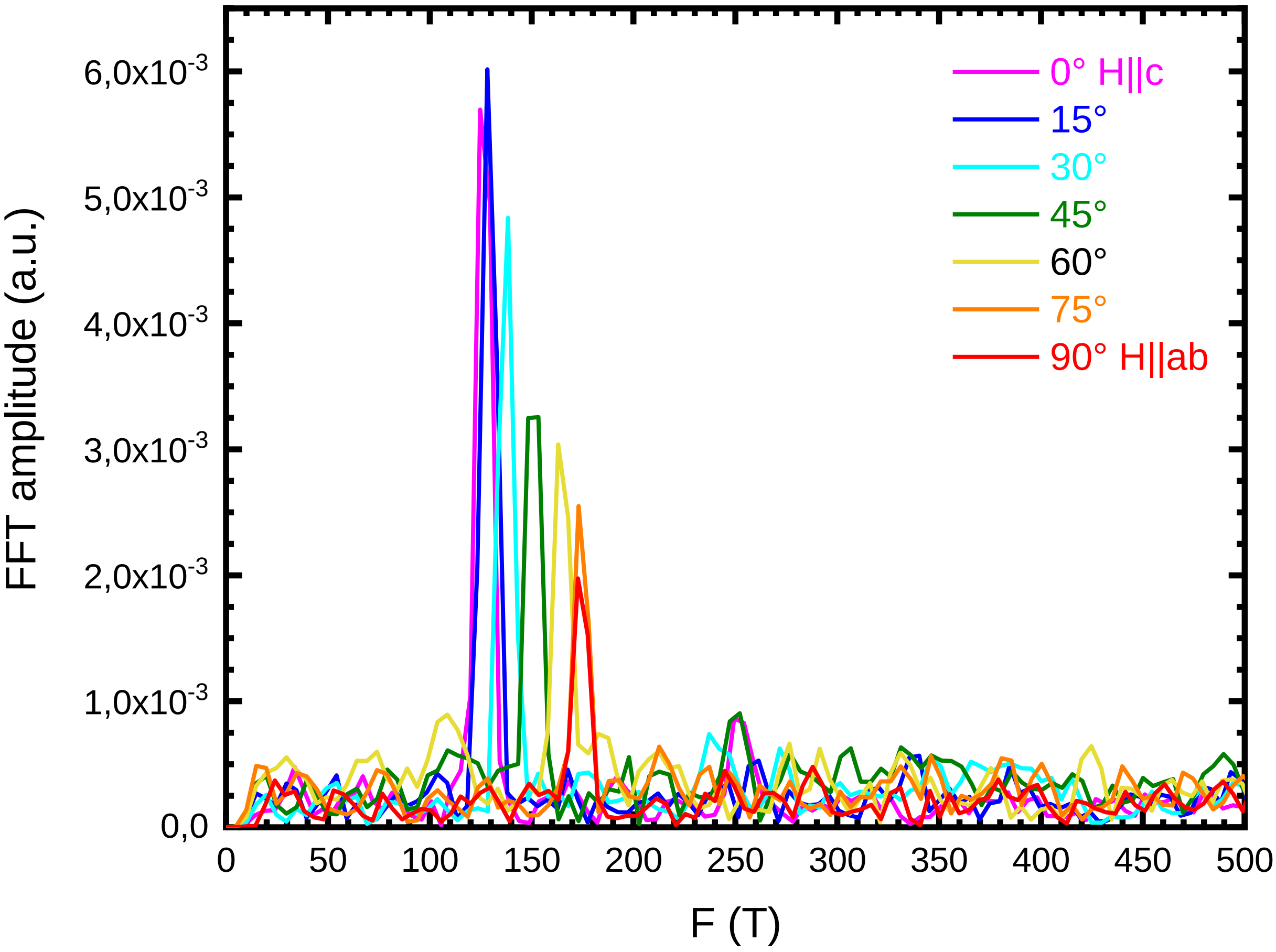

Fig. 8. The FFT spectra of the $C_{osc}/T$ data when the external magnetic field is oriented towards the *c* axis at different angles (0.73 K).

In Fig. 8, we show the FFT spectra of the MQOs when the field is oriented towards the *c*-axis at different angles at 0.73 K, allowing us to determine the angular dependence of the main frequency. The total phase and consequently the Berry phase are preserved at different angles. The fit of $C_{osc}(T = 0.73$ K, $B)$ at $\mu_0 H \parallel ab$ is shown in Fig. 9.

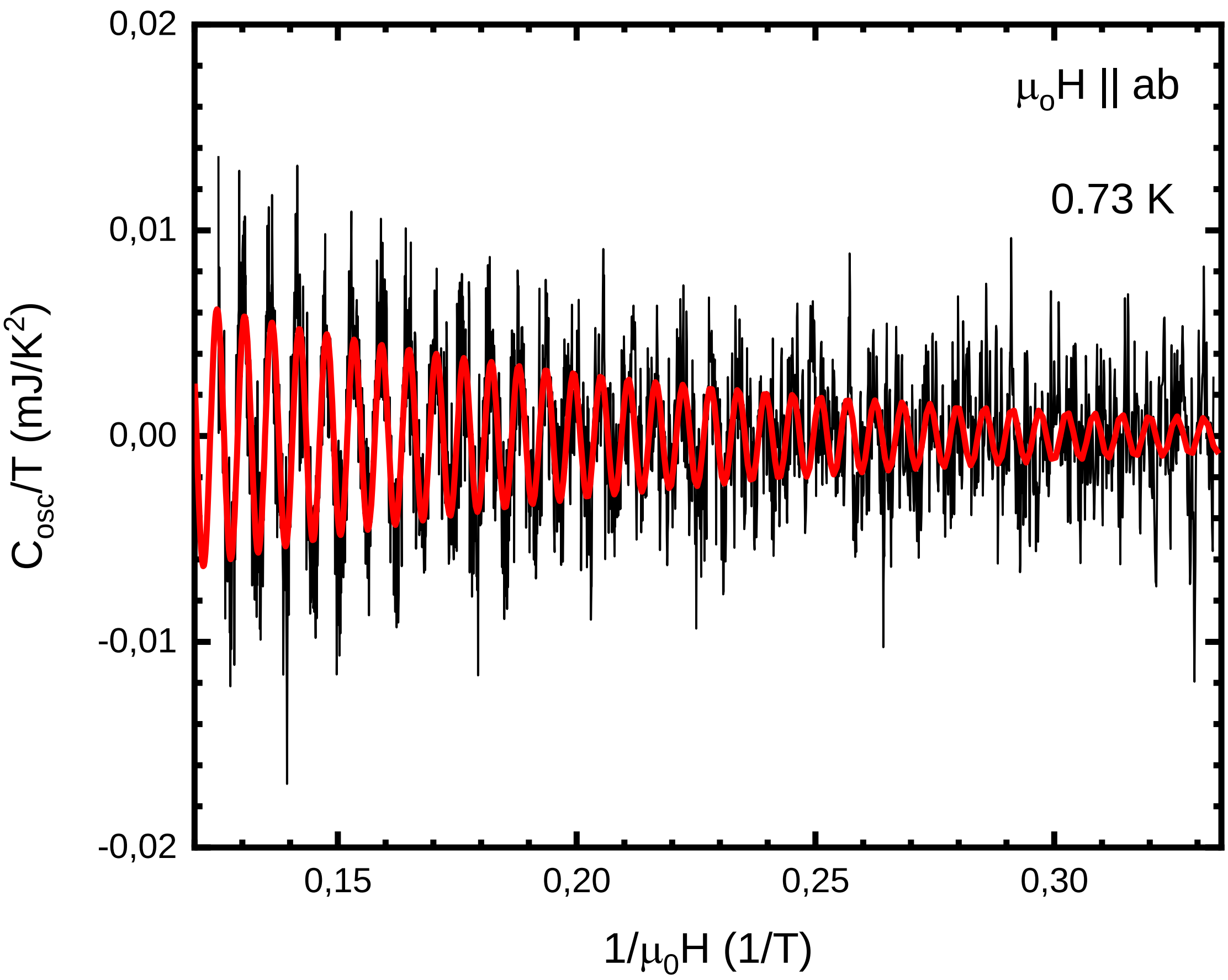

Fig. 9. Experimental $C_{osc}/T$ (black curve, $\mu_0$H || ab) directly fitted (red curve) to the LK model with the only main harmonic of 174.2 T at 0.73 K.

As can be seen, the amplitude of the signal decreased almost threefold in comparison with that of $C_{osc}(T = 0.73$ K, $B)$ at $\mu_0 H$ || c, while the main frequency increased to 174.2 T. The increase in frequency with angle indicates that the extremal Fermi-surface cross-section increases when the magnetic field is rotated, reflecting anisotropy in the electronic structure.

Fig. 10 summarizes the result showing the angular dependences of the FFT amplitude (red points) and frequency (blue points). The angular dependence of the frequency is compared with that calculated de Haas-van Alphen frequencies (green curve) using the SKEAF code [22], confirming the contribution from the γ pocket only.

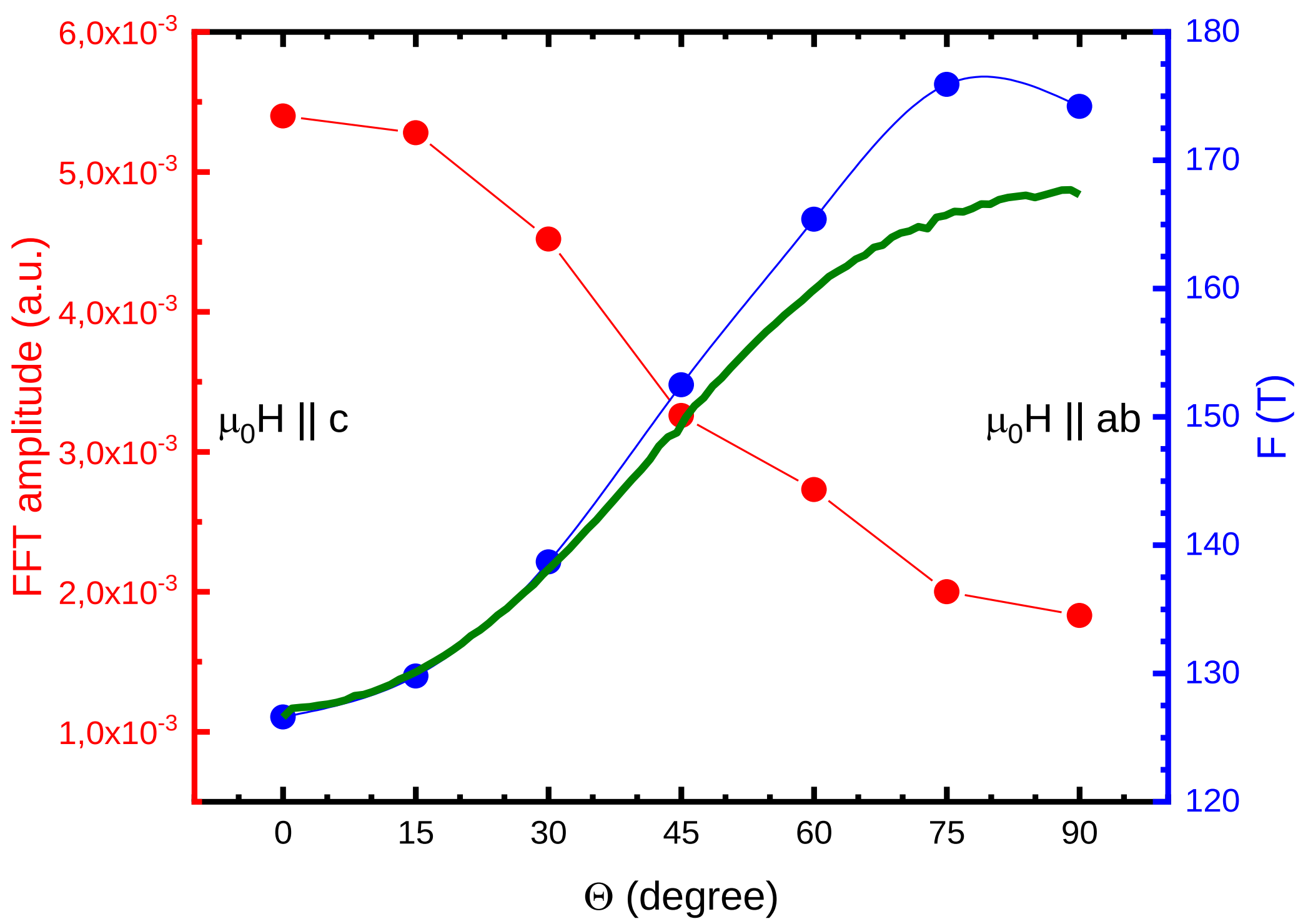

Fig. 10. The angular dependences of the FFT amplitude (red dots) and frequency (blue dots). Angle increases from 0 for $\mu_0 H \parallel c$ to 90 degrees for $\mu_0 H \parallel ab$. Calculated angular-dependent frequencies (green curve) corresponding to the extremal cross-section of the γ pocket.

**Discussion**

According to previous transport studies, $V_2Ga_5$ is a multiband metal containing both electron- and hole-like carriers. Nevertheless, its low-temperature Hall response is dominated by holes, indicating that hole pockets make the principal contribution to the charge transport properties. In this context, the observed quantum-oscillation branch with a frequency of 126.6 T can be naturally associated with the hole-like Fermi-surface sheet. However, this assignment is consistent with the band-structure analysis [6, 22], and the lowest frequencies correspond to the extremal cross-section of the γ Fermi pocket.

The oscillation frequency of 126.6 obtained in the present work is in excellent agreement with the value reported from magnetization measurements by [5] and [22]. Furthermore, both the FFT analysis and the direct fitting of the oscillatory component yield the same $m^* = 0.245m_e$, demonstrating the internal consistency and reliability of the calorimetric analysis. The obtained effective mass lies between the values previously reported from de Haas–van Alphen measurements, namely $m^* = 0.159m_e$ in [5] and $m^* = 0.31m_e$ in [22]. Considering the different experimental techniques and analysis procedures employed, the overall agreement is satisfactory and supports the validity of the present results.

An additional important result concerns the Berry phase. The analysis yields a Berry phase of ~0.8π, taking $\delta = +\pi/4$. The value is close to the phase of ~0.9π reported by [5] and significantly deviates from the trivial value expected for conventional parabolic bands. Such a nontrivial Berry phase is consistent with the presence of topologically nontrivial electronic states participating in the quantum oscillations. The agreement between the present thermodynamic measurements and previous magnetization studies therefore supports the unconventional character of the electronic structure of $V_2Ga_5$.

The γ pocket exhibits an elliptical cross-section centered at the *Z* point, with an integrated orbital composition, determined by band-resolved projected density of states, of 57 % V *d*-orbitals present at the poles and 43 % Ga *p*-orbitals mostly located at the equator of the pocket. We parameterize the cyclotron orbit by the cyclotron phase angle $\theta$, tracing the constant-energy extremal contour in *k*-space. The evolution of the Bloch wave function

$$|u(\theta)\rangle = \alpha(\theta)|V\rangle + \beta(\theta)e^{i\zeta(\theta)}|Ga\rangle \tag{8}$$

captures the hybridization of the V and Ga states. Evaluating the Berry phase via

$$\Phi_B = \oint \langle u(\theta)|i\partial_\theta|u(\theta)\rangle d\theta \,, \tag{9}$$

the non-zero contribution arises from the orbital weighted phase twist:

$$\Phi_B = \langle|\beta|^2\rangle \Delta\zeta \,, \tag{10}$$

where $\langle|\beta|^2\rangle$ represents the integrated Ga *p*-orbital weight and $\Delta\zeta$ is the total relative phase twist acquired by the Ga component relative to the V component over one closed orbit. The experimental observation of an isotropic $\Phi_B \approx 0.8\pi$ signifies that the net Berry flux is an invariant of the γ pocket. Using the integrated upper limit of $\langle|\beta|^2\rangle \approx 0.43$, we find $\Delta\zeta = \Phi_B / \langle|\beta|^2\rangle \approx 0.8\pi / 0.43 \approx 1.86\pi$, which is close to $2\pi$ topological phase winding. Because the Berry phase reflects this $2\pi$ topological winding of the hybridization phase $\Delta\zeta$ across the Brillouin zone, the phase accumulation remains independent of the specific cyclotron orbit geometry or field orientation. Thus, $\Phi_B$ serves as a macroscopic probe of the intrinsic parity-dependent orbital hybridization, with the observed value acting as a geometric projection of this $2\pi$ winding scaled by the specific Ga orbital fraction of the hybridized band.

**Conclusions**

In summary, we have investigated the magneto-quantum oscillations in single-crystalline $V_2Ga_5$ using heat-capacity measurements performed via highly sensitive ac calorimetry. A dominant oscillation frequency of 126.6 T and an effective cyclotron mass of $0.245m_e$ were consistently obtained from both FFT analysis and direct fitting of the experimental data, confirming the reliability of the bulk γ pocket parameters. For the first time in $V_2Ga_5$, a Dingle temperature of 1.25 K was evaluated from calorimetric quantum oscillation measurements, enabling the determination of the quantum relaxation time and electronic mean free path, which provide critical insight into carrier scattering processes and sample quality. The angular dependence of the magneto-quantum oscillations is supported by first-principles calculations of the cyclotron frequencies originating from the extremal cross-sectional areas of the γ Fermi pocket. Crucially, we demonstrate that the net Berry flux is invariant with respect to the magnetic field orientation, a direct consequence of a conserved $2\pi$ hybridization phase twist within the γ pocket. These results highlight the effectiveness of ac calorimetry for uncovering subtle geometric phase phenomena and provide valuable thermodynamic insights into the complex electronic structure of intermetallic systems.

**Acknowledgments**

This work was supported by the Slovak Research and Development Agency under the Contracts No. APVV-23-0624, the Science Grant Agency Projects No. VEGA 2/0073/24, No. VEGA 1/0472/25, and No. VEGA 1/0104/25, the Slovak Academy of Sciences Projects No. IMPULZ IM-2021-42. Research performed at Gdansk University of Technology was supported by the National Science Center (Poland), Project No. 2022/45/B/ST5/03916.